\documentclass[preprint,3p,12pt]{elsarticle}

\usepackage{amssymb}
\usepackage{amsmath}
\usepackage[hidelinks]{hyperref}
\usepackage{color}
\usepackage{ulem}

\newcommand{\beq}{\begin{eqnarray}}
\newcommand{\eeq}{\end{eqnarray}}

\begin{document}
\begin{frontmatter}

\title{Using net quark number gain to probe the phases of QCD}

\author[CPhT]{V. Tomas Mari Surkau\corref{c1}}
\ead{victor-tomas.mari-surkau@polytechnique.edu}

\author[CPhT]{Urko Reinosa}
\ead{urko.reinosa@polytechnique.edu}

\cortext[c1]{Corresponding author}

\affiliation[CPhT]{organization={Centre de Physique Th\unexpanded{\'e}orique, CNRS, \unexpanded{\'E}cole Polytechnique, IP Paris, F-91128 Palaiseau, France.}}

\begin{abstract}
We discuss an observable that probes the content of a QCD medium at finite temperature and chemical potential, the net quark number gain. It is the response of the thermal bath to a static quark or antiquark probe. While insignificant at high temperatures, it reveals the bath's tendency to form meson-like or baryon-like configurations (depending on the probe and chemical potential) at low temperatures, which would screen the probe’s color charge. The net quark number gain also helps explain how a single quark/antiquark can be added to a supposedly confining medium in the first place: the latter provides the missing quarks/antiquarks to form hadron-like states. We sketch the derivation of this general result for temperatures much smaller than the constituent quark masses and discuss possible further applications to study the various features of the QCD phase diagram.
\end{abstract}


\end{frontmatter}

\section{Introduction}\label{sec: Intro}
Nuclear physics experiments observe excitations called hadrons, which are bound states of quarks. This stands in stark contrast to the elementary degrees of freedom in Quantum Chromodynamics (QCD), the theory that has successfully predicted the outcomes of these experiments. Its fundamental degrees of freedom are quarks and gluons, charged under the SU(3) color group, that are \textit{a priori} not necessarily bound into hadrons. However, their color charge has not been observed in experiments, where we only find color-neutral hadrons into which the quarks and gluons are confined. Lattice simulations observe a rapid change in the degrees of freedom as a nuclear medium is heated, which is understood as a crossover from a gas of hadrons to a deconfined quark-gluon phase \cite{HotQCD2014EosQCD}. This is corroborated by heavy-ion experiments at sites like CERN and RHIC, where a Hadron Resonance Gas model accurately describes the chemical freeze-out yields of heavy-ion collisions \cite{Andronic2018HRGFreeze-out}, but signatures such as anisotropic flow \cite{Luzum2008FlowRHIC} and jet quenching \cite{STAR2003JetQuenching} indicate the presence of a strongly coupled, locally thermalized quark-gluon plasma during the intermediate stages of these collisions.
Many experimental and theoretical efforts are being made to better understand these extreme states of matter and to draw a comprehensive QCD phase diagram. For this, we need to determine the active degrees of freedom at different temperatures and chemical potentials. Lattice simulations suffer from the sign problem at finite chemical potentials, prompting alternatives like functional QCD and model studies to establish our predictions of the phases in the higher-density regions.

As the transition in QCD is actually a crossover, at least for sufficiently small chemical potentials, there is no clear-cut order parameter; instead, many observables sensitive to the degrees of freedom evolve rapidly through the temperature window around the crossover. \cite{Aarts2023PhasePhysics, Aoki2006ThePhysics} However, in certain limits of the theory, it turns into a true phase transition with strict order parameters allowing clean interpretations. In the chiral limit, the quark condensate acts as an order parameter and, via spontaneous chiral symmetry breaking, gives rise to large hadron masses at low temperatures. In the opposite limit of infinitely heavy quark masses, the Polyakov loop is an order parameter for the confinement of static color charges in the (anti-)fundamental representations, i.e., (anti-)quarks. It is the exponential of the medium's free energy difference without and with an additional color charge, and the center symmetry of Yang-Mills theory forces it to vanish, corresponding to an infinite free energy difference. This is interpreted as it being forbidden to add static quarks or antiquarks to the thermal bath. At large enough temperatures, this symmetry also breaks spontaneously, and adding isolated color charges becomes possible at a finite energy cost \cite{Svetitsky1982CriticalTransitions, Polyakov1978ThermalLiberation}.

As visualized by the Columbia plot, remnants of these transitions persist away from the exact limits, and the order parameters keep their characteristic behavior, functioning as indicators for the phase transition. For the chiral condensate, the interpretation remains simple: the symmetry is significantly broken at low energies, in addition to the explicit breaking by the free quark masses, resulting in the increased hadron masses. The situation is more delicate for the Polyakov loops, which are small but non-zero in the low-temperature phase and grow towards unity around the transition. The explicit breaking of center symmetry by the quarks, even if they are heavy, allows, at least in principle, the existence of an isolated quark in a low-temperature QCD medium, since the free energy cost, albeit large, is not infinite anymore. This contradicts the experimental observations and theoretical expectation that the quarks are confined into hadrons, the low-temperature degrees of freedom.

In a recent work \cite{MariSurkau2025MesonsBaryons}, we reconciled this apparent contradiction by correlating the Polyakov loop with the net quark number of the medium. This gives access to the medium's net quark number gain upon the addition of a single static quark or antiquark, which is the difference of the medium's net quark number with and without the color source, plus the net quark number of the probe. We expect the medium to bring forth the necessary quarks or antiquarks to screen the color charge of the source into a hadron, at least below the transition temperature, which should then be reflected in the medium's changed net quark number. To access it, we use the Polyakov loop potential, the extremum of which gives the Polyakov and anti-Polyakov loops at different temperatures and chemical potentials.

\section{Polyakov loops}\label{sec: pol loop}
The Polyakov loop $\ell$ and anti-Polyakov loop $\bar\ell$ are related to the free energy differences from adding a static quark $\Delta F_q$ or antiquark $\Delta F_{\bar{q}}$ as
\begin{equation}\label{eq: l lbar}
    \ell=e^{-\beta \Delta F_q},\quad \bar\ell=e^{-\beta \Delta F_{\bar{q}}},
\end{equation}
where $\beta=1/T$ is the inverse temperature. The derivative of the free energy with respect to the quark chemical potential $\mu$ corresponds to the associated charge, in this case, the net quark number $Q=-\partial_\mu F$. Therefore, the response of the medium to the quark/antiquark probe in terms of net quark number is $\Delta Q_{q,\bar{q}}=-\partial_\mu\Delta F_{q,\bar{q}}$. Using the relation \eqref{eq: l lbar}, the medium's net quark number gain upon adding a quark or antiquark is then
\begin{equation}\label{eq: Delta Q def}
    \Delta Q_q+1=1+T\partial_\mu\ln\ell, \quad \Delta Q_{\bar{q}}-1=-1+T\partial_\mu\ln\bar\ell,
\end{equation}
where we accounted for the quark number of the probe $\pm1$ in addition to the medium's response. Note that this is a gauge- and RG-invariant quantity, and hence a proper theoretical observable. To gain access to the net quark number gain, we thus only need to know the $\mu$-dependence of the Polyakov loops. To access them, we need the thermodynamic potential $\omega=\omega(T,\mu,\ell,\bar\ell,\langle\bar{q}q\rangle,\dots)$, which generally depends on all the observables of the theory and on the external thermodynamic parameters such as $T$ and $\mu$. The physical solutions $\ell(T,\mu)$ and $\bar\ell(T,\mu)$ minimize $\omega$ at any given $T$ and $\mu$.\footnote{Note that in applications the solutions are usually saddle points of the potential instead, seen as a remnant of the sign problem in the continuum.}

Note that a $\mu$-derivative of the potential gives access to the net quark number density $n_q=-\partial_\mu\omega$, and the order parameters can therefore feed back into $n_q$. However, it corresponds to the net quark number density of the system without an added quark/antiquark, in contrast to the net quark number gain, which is a global observable rather than a density and is the total difference of net quark number in a system with the added quark/antiquark to the system without. The two quantities encode fundamentally different information.

Accessing the full thermodynamic potential in QCD is prohibitively hard, so it is generally approximated. This is most simply achieved in the case of heavy quark QCD, where the Polyakov loops function best as order parameters for the QCD transition. We work in this setup, but we argue below that for sufficiently small temperatures, the results should also hold in real QCD. 
Specifically, one can separate a pure gauge and a quark part $\omega\simeq V_{\rm glue}+V_{q}$ of the thermodynamic potential. For the pure gauge part, we require that it must respect the center symmetry of Yang-Mills theory, i.e. 
\begin{equation}\label{eq: center sym}
    V_{\rm glue}(\ell,\bar\ell)=V_{\rm glue}(e^{i2\pi/3}\ell,e^{-i2\pi/3}\bar\ell),
\end{equation}
and that it is confining at low $T$, i.e., the extrema obey $\ell,\bar\ell\xrightarrow{T\to0}0$, as seen in lattice simulations. \cite{McLerran:1981pb, Brown:1988qe}
Lastly, we assume that $V_{\rm glue}$ has a $T$-power-law behavior at low $T$. This behavior is common to all model potentials used in the literature \cite{Fukushima2004ChiralLoop, Ratti:2005jh, Lo:2013hla, Reinosa:2014ooa, MariavanEgmond2022ATemperature, MariSurkau2024Deconfinement}
and is supported by the observation that confinement is related to massless modes in the gauge potential \cite{Gupta:2007ax, Braun:2007bx}. Apart from these three conditions, our reasoning is fully model-independent concerning the exact form of the glue potential. 
The quark potential is given by the quark-loop contribution \cite{Fukushima2004ChiralLoop}
\beq
    V_{q}(\ell,\bar\ell,T,\mu)=-\frac{TN_f}{\pi^2}\int_0^\infty dq\,q^2&&\kern-.6cm\Big\{\ln\Big[1\!+\!3\ell e^{-\beta(\varepsilon_q-\mu)}\!+\!3\bar\ell e^{-2\beta(\varepsilon_q-\mu)}\!+\!e^{-3\beta(\varepsilon_q-\mu)}\Big] \nonumber\\
    && +\ln\Big[1\!+\!3\bar\ell e^{-\beta(\varepsilon_q+\mu)}\!+\!3\ell e^{-2\beta(\varepsilon_q+\mu)}\!+\!e^{-3\beta(\varepsilon_q+\mu)}\Big]\Big\},
\eeq
with $\varepsilon_q=\sqrt{q^2+M^2}$ and $N_f$ the number of quark flavors, which for simplicity we assume to be degenerate. As a consequence of the power-law behavior of $V_{\rm glue}$, the exponentially suppressed quark part is less relevant at low $T$ (and $\mu<M$), and the extrema will still be at the confining point given by the glue part. 

This one-loop approximation is most accurate in the case of heavy quark masses $M\gg T$, which effectively suppress higher contributions. However, we expect the results to extend to physical QCD, since using the well-tested expansion in the inverse number of colors in the Landau gauge, with the fact that the pure glue coupling is not that large \cite{Duarte:2016iko, Reinosa:2017qtf}, the quark contribution to the Polyakov loop potential is given by an effective one-loop contribution involving the rainbow-resummed quark propagator. At low temperatures, it should be dominated by low momenta and thus by the large constituent quark mass $M_{\rm const}$ given by chiral symmetry breaking, so we expect similar results in the regime $T\ll M_{\rm const}$.

When $T\ll M$, $|\mu|<M$, and $\ell,\bar\ell\ll1$ we can approximate
\begin{equation}\label{eq: Vq approx}
    V_q\simeq-\big\{\ell (e^{\beta\mu}f_{\beta M}+e^{-2\beta\mu}f_{2\beta M})+\bar\ell (e^{2\beta\mu}f_{2\beta M}+ e^{-\beta\mu}f_{\beta M})+(e^{3\beta\mu}+e^{-3\beta\mu})f_{3\beta M}/3\big\},
\end{equation}
where we defined
\begin{equation}\label{eq: f_y}
    f_y=\frac{3N_fTM^3}{\pi^2}\int_0^\infty dx\,x^2e^{-y\sqrt{1+x^2}}\sim \frac{9N_fTM^3}{\sqrt{2}\pi^{3/2}}e^{-y}y^{-3/2},
\end{equation}
and the asymptotic expression on the right holds for $y\gg1$. The Polyakov and anti-Polyakov loops are found by extremizing the potential, which, using Eq.~\eqref{eq: Vq approx}, gives
\beq
    0\simeq\partial_\ell V_{\rm glue}-(e^{\beta\mu}f_{\beta M}+e^{-2\beta\mu}f_{2\beta M}), \quad 0\simeq\partial_{\bar\ell} V_{\rm glue}-(e^{-\beta\mu}f_{\beta M}+e^{2\beta\mu}f_{2\beta M}).
\eeq
Since at low $T$ the Polyakov loops approach $\ell=\bar\ell=0$, we linearize around this point.
Using that by the center-symmetry \eqref{eq: center sym} $\partial_\ell V_{\rm glue}=\partial_{\bar\ell} V_{\rm glue}=\partial_{\ell}\partial_{\ell} V_{\rm glue}=\partial_{\bar\ell}\partial_{\bar\ell} V_{\rm glue}=0$ when the derivatives are evaluated at $\ell,\bar\ell=0$ the linearized equations are readily solved to get
\begin{equation}\label{eq: pol loops results}
    \ell\simeq\frac{1}{\partial_\ell\partial_{\bar\ell}V_{\rm glue}|_{\ell,\bar\ell=0}}(e^{-\beta\mu}f_{\beta M}+e^{2\beta\mu}f_{2\beta M}),\quad
    \bar\ell\simeq\frac{1}{\partial_\ell\partial_{\bar\ell}V_{\rm glue}|_{\ell,\bar\ell=0}}(e^{\beta\mu}f_{\beta M}+e^{-2\beta\mu}f_{2\beta M}).
\end{equation}
The Polyakov loop can then be calculated by using a specific form of $V_{\rm glue}$. We note that the cases $|\mu|\geq M$ and of higher $T$ are not included in the approximation \eqref{eq: Vq approx}, and the results are therefore different in those regimes. At high temperatures, the Polyakov loops saturate towards 1, and at high chemical potentials, the quark part is no longer exponentially suppressed, and the dynamics at low temperatures depend on the exact scaling of $V_{\rm glue}$. However, for heavy quarks, this regime is always interpreted as deconfined. In either case, the Polyakov loops will not have the exponential scaling in $\mu$ anymore, which becomes crucial through the derivative of their logarithm in \eqref{eq: Delta Q def}, which gives the net quark number gain. For a detailed treatment of the various cases, see \cite{MariSurkau2025MesonsBaryons}.

\begin{figure}
    \centering
    \includegraphics[scale=.605]{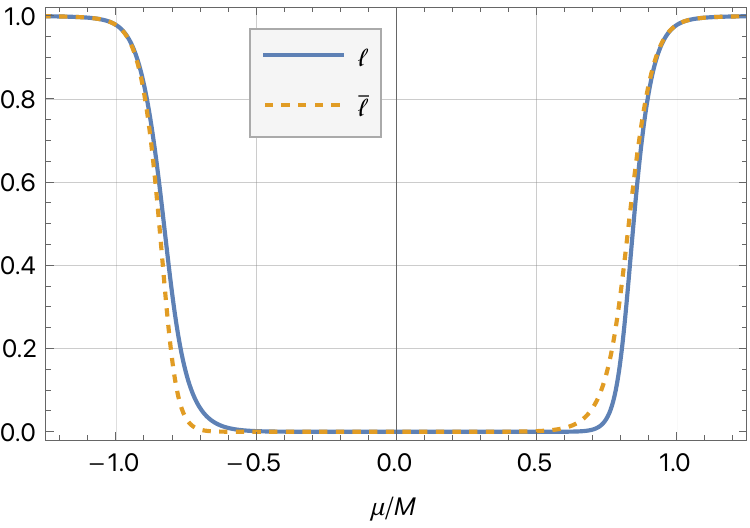}\hfill
    \includegraphics[scale=.63]{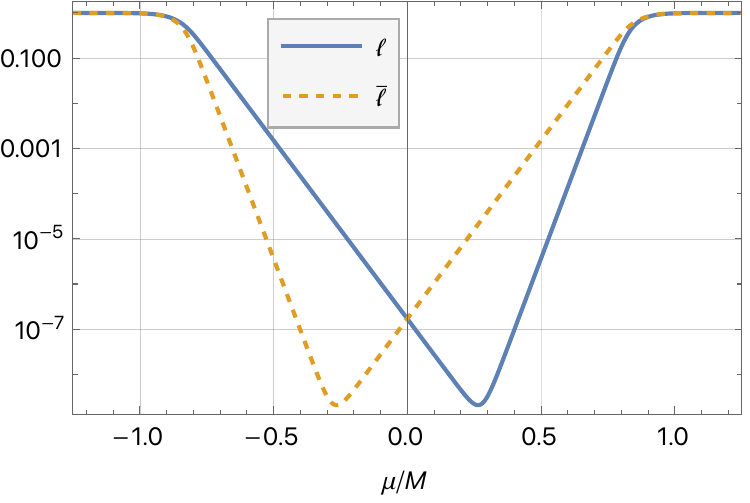}
    \caption{Polyakov loop $\ell$ (solid) and anti-Polyakov loop $\bar\ell$ (dashed) at a low temperature $T\ll M$ as a function of chemical potential $\mu$ normalized to the quark mass $M$, in (\emph{left}) a linear scale and (\emph{right}) a logarithmic scale. The logarithmic scale reveals a striking behavior of differing but constant slopes in the confined region, invisible in the linear scale plot due to the Polaykov loops' exponential suppression.}
    \label{fig: Pol loops linear log}
\end{figure}

As $V_{\rm glue}$ is $\mu$-independent, the Polyakov loops' $\mu$-dependence is now fully known. There are two regimes in $\mu$ with different behaviors, depending on which combination of exponentials dominates. The Polyakov loop $\ell$ is monotonically decreasing for $\mu\lesssim M/3$ and increasing after, while the anti-Polyakov loop $\bar\ell$ is decreasing for $\mu\lesssim-M/3$ and increasing after, as expected from charge conjugation.\footnote{Note that \eqref{eq: pol loops results} holds for $T\to0$, but is not fully consistent at finite $T$. Due to the different exponential scalings of $\ell$ and $\bar\ell$, one can be of the order of the square of the other. A consistent expansion does not change the result up to a small shift of $\mu$ at finite temperatures \cite{MariSurkau2025MesonsBaryons}.} This behavior is clearly seen in the logarithmic plot to the right of Fig.~\ref{fig: Pol loops linear log}. This plot also magnifies the different exponential suppression of the Polyakov loops in the confined phase, with characteristic linear dependencies of $\ln\ell$ and $\ln\bar\ell$ in $\mu$. As we discuss below, these linear regimes have a neat and simple physical interpretation. To our knowledge, these behaviors had never been pointed out before. We advocate for plotting the logarithm of Polyakov loops in other applications as well, to make their behavior and interpretation clearer.

The results in Fig.~\ref{fig: Pol loops linear log} were calculated using the center-symmetric Curci-Ferrari model \cite{MariavanEgmond2022ATemperature, MariSurkau2024Deconfinement} for $V_{\rm glue}$ and with $M=2\,$GeV. The same behavior is seen for any temperature almost up to $T_c$. Analogous results are obtained with the other popular models for the Polyakov loop potential. The interpretation of this characteristic behavior comes from analyzing the slope in the logarithmic plot, which, as per Eq.~\eqref{eq: Delta Q def}, corresponds to the net quark number gain up to a factor of $T$.

\section{Net quark number gain}\label{sec: DeltaQ}
\begin{figure}
    \centering
    \includegraphics[scale=.6]{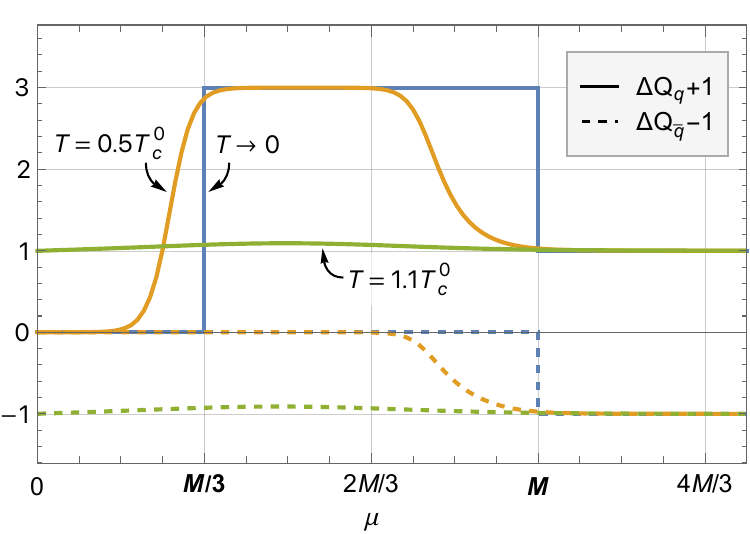}\hfill
    \includegraphics[scale=.63]{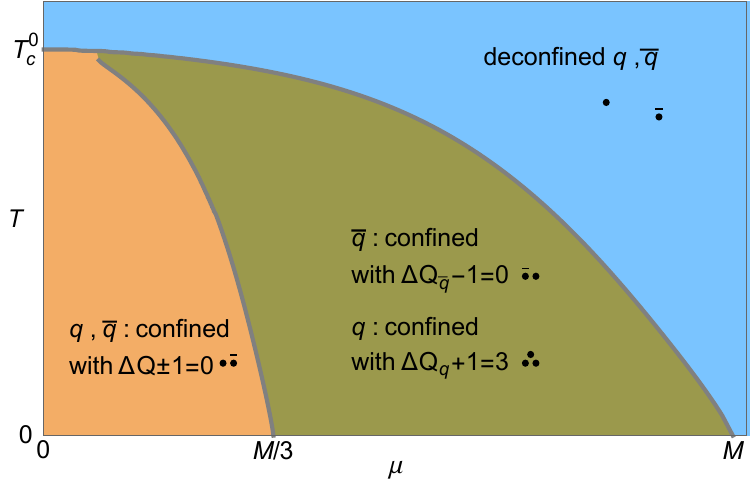}
    \caption{\emph{Left:} Net quark number gain upon addition of a quark (solid) or an antiquark (dashed) as a function of $\mu$ in units of the quark mass $M$, for a temperature below and above the transition temperature as well as in the zero temperature limit, where it becomes a step function.\\
    \emph{Right:} Phase diagram of heavy quark QCD characterized by the net quark number gain. In addition to the separation of the deconfined (upper) and confined (both lower) phases, we distinguish where quarks are screened into meson-like (0, lower left) and baryon-like (3, lower-right) configurations.}
    \label{fig: Delta Q}
\end{figure}

Using the results \eqref{eq: pol loops results} in Eq.~\eqref{eq: Delta Q def}, the model-dependent prefactor drops out, and we get
\begin{equation}\label{eq: Delta Q result}
    \Delta Q_q+1\simeq\frac{3}{1+e^{-3\beta\mu}f_{\beta M}/f_{2\beta M}}, \quad \Delta Q_{\bar{q}}-1\simeq\frac{-3}{1+e^{3\beta\mu}f_{\beta M}/f_{2\beta M}}.
\end{equation}
In the limit $T\to0$ these quantities becomes step functions, with $\Delta Q_q+1\simeq0$ for $\mu<M/3$ and $\Delta Q_q+1\simeq3$ for $\mu>M/3$, and similarly $\Delta Q_{\bar q}-1\simeq-3$ for $\mu<-M/3$ and $\Delta Q_{\bar q}-1\simeq0$ for $\mu>-M/3$, as visualized in Fig.~\ref{fig: Delta Q}. This resolves the supposed contradiction from above: An added color source does not exist in isolation; instead, the medium arranges to produce the appropriate number of quarks or antiquarks, thereby forming a hadron-like state.

Interestingly, depending on the value of the chemical potential, corresponding to the surplus of quarks or antiquarks (baryons or anti-baryons) in the medium, different solutions are favored. At large negative chemical potentials, the quarks get screened into a meson-like state characterized by a net quark number gain $\Delta Q_q+1\simeq0$ and antiquarks into an anti-baryon-like state, characterized by $\Delta Q_{\bar q}-1\simeq-3$. Closer to net quark number equilibrium, $|\mu|<M/3$, both quarks and antiquarks lead to meson-like net quark number gains, with the medium's average response just being a change of net quark number opposite to the introduced charge giving $\Delta Q_{q,\bar{q}}\pm1\simeq0$. For further increasing $\mu$, an antiquark will still prompt a medium response of just an additional quark, while the net response to a quark is two quarks, leading to a baryon-like state with $\Delta Q_q+1\simeq3$. Instead, in the deconfined phase, a quark or antiquark probe can be added to the medium without a significant response in terms of net quark number, as reflected by $\Delta Q_{q,\bar q}\simeq0$. The net quark number gain is then just the quark number of the probe $\pm1$. This allows us to draw an extended QCD phase diagram separating these regions, see Fig.~\ref{fig: Delta Q}.

The above result can also be explained from clear thermodynamic arguments in a simplified non-relativistic, heavy quark case. In the $ T\to 0$ limit, the dominant state minimizes the exponent $H-\mu Q$ in the partition function, which can be approximated as $H-\mu Q \approx (N_q+N_{\bar q})M-\mu (N_q-N_{\bar q})$, where $N_q,\, N_{\bar q}$ are the numbers of quarks/antiquarks. If there are only hadronic degrees of freedom, then upon adding a quark we must have $N_q-N_{\bar q}+1=0\mod{3}$, i.e., $(N_q, N_{\bar{q}})=(0,1),\,(2,0)\,(3,1),...$. Then the values of $(N_q,N_{\bar q})$ that minimize $H-\mu Q$ are (0,1), i.e. meson-like, for $\mu<M/3$ and (2,0), i.e. baryon-like, for $\mu>M/3$.

These arguments and the formula of Eq.~\eqref{eq: Delta Q result} are valid in the $T\to0$ limit, but at finite temperatures, some care needs to be taken, see also the discussion in \cite{MariSurkau2025MesonsBaryons}. In general, both the meson-like and baryon-like states now have a non-zero weight in the partition function, and the net quark number gain, which in the grand canonical ensemble is only determined on average, will be influenced by both types of states. This explains the smooth transition behavior between 0 and 3 seen in Fig.~\ref{fig: Delta Q} and the shift of the position of the transition from meson- to baryon-like states towards smaller $\mu$ as $T$ increases. The exact value of $\mu$ depends slightly on the model used for the Polyakov loop potential, while the existence of the transition and the $T\to0$ value of $M/3$ are universal.

\section{Conclusion and Outlook}
We reviewed a theoretical observable, the net quark number gain, which appears sensitive to the net quark number content of a QCD medium's active degrees of freedom. We showed this for heavy quarks and argued that the approximations hold up in real QCD at low enough temperatures. We confirmed this expectation in a model calculation, coupling a Polyakov loop potential to an NJL model; the results will be shown elsewhere. In contrast to the heavy quark case, the plateaus smooth out already before the transition, as chiral symmetry breaking weakens and thermal fluctuations increase.

Furthermore, the net quark number gain could open new perspectives on the QCD phase diagram. For example, at large $\mu$ color superconducting phases are expected to appear, with different dominant degrees of freedom. In this case, it would be interesting to verify whether the net quark number gain could take the value 2, corresponding to the diquarks' net quark number content. In the medium temperature regime, it would be interesting to study whether it is sensitive to the persistence of chirally symmetric hadronic states above the chiral crossover \cite{Glozman2022ChiralDiagram}. When it comes to studying the structure of the phase diagram, we also note that near a critical point, the net quark number gain will diverge, as the slope of the Polyakov loops becomes unbounded. 

As the net quark number gain is essentially the correlation of the Polyakov loop operator $\Phi$ with the net quark number $\langle\Phi Q\rangle$, other combinations of Polyakov loops and conserved charges could reveal interesting properties about QCD media as well. A clear candidate is the color Casimir $\langle \Phi Q^aQ^a\rangle$, which could give access to the color representation formed by the medium in response to the probe, and should confirm that in the hadronic phase, a neutral object can be formed with the probe. As neutron stars have a finite isospin density, correlating the Polyakov loops with the isospin might also give interesting insights for those systems.

\bibliographystyle{elsarticle-num}
\bibliography{references}
\end{document}